\documentclass{llncs}

\usepackage[utf8]{inputenc}
\usepackage{graphicx}
\usepackage[boxed,vlined,ruled,linesnumbered]{algorithm2e}
\usepackage{cite}
\usepackage{amsmath}
\usepackage{amsfonts}
 
\usepackage{color}
\usepackage{multicol}
\usepackage{multirow}
\usepackage{colortbl}
\usepackage{enumerate}
\usepackage{tikz}
\usetikzlibrary{shadows}
\usetikzlibrary{positioning}
\usetikzlibrary{fit}
\usetikzlibrary{backgrounds}
\usetikzlibrary{chains}
\usetikzlibrary{matrix}
\usetikzlibrary{snakes}
\usetikzlibrary{shapes}
\usetikzlibrary{calc}
\usetikzlibrary{arrows}

\usepackage{longtable}

\newcommand{\scsc}{\textsf{SC$^\mathsf{2}$}\xspace}

\newcommand{\smtlib}{\textit{SMT-LIB}\xspace}

\newcommand{\smtrat}{{\small\texttt{SMT-RAT}}\xspace}

\newcommand{\altergo}{{\small\texttt{Alt-Ergo}}\xspace}
\newcommand{\opensmt}{{\small\texttt{OpenSMT2}}\xspace}
\newcommand{\cvc}{{\small\texttt{CVC4}}\xspace}
\newcommand{\yices}{{\small\texttt{Yices2}}\xspace}
\newcommand{\mathsat}{{\small\texttt{MathSAT}}\xspace}

\newcommand{\minismt}{{\small\texttt{MiniSmt}}\xspace}
\newcommand{\isat}{{\small\texttt{iSAT3}}\xspace}
\newcommand{\zthree}{{\small\texttt{Z3}}\xspace}

\newcommand{\verit}{{\small\texttt{veriT}}\xspace}
\newcommand{\aprove}{{\small\texttt{AProVE}}\xspace}

\newcommand{\maple}{{\small\texttt{Maple}}\xspace}

\newcommand{\reduce}{{\small\texttt{Reduce}}\xspace}
\newcommand{\redlog}{{\small\texttt{Redlog}}\xspace}
\newcommand{\cocoa}{{\small\texttt{CoCoA}}\xspace}

\newcommand{\cocoalib}{{\small\texttt{CoCoALib}}\xspace}
\newcommand{\singular}{{\small\texttt{Singular}}\xspace}

\newcommand{\macsyma}{{\small\texttt{Macsyma}}\xspace}
\newcommand{\sac}{{\small\texttt{SAC}}\xspace}
\newcommand{\scratchpad}{{\small\texttt{SCRATCHPAD}}\xspace}
\newcommand{\axiom}{{\small\texttt{AXIOM}}\xspace}
\newcommand{\magma}{{\small\texttt{MAGMA}}\xspace}

\newcommand{\mathematica}{{\small\texttt{Mathematica}}\xspace}

\newcommand{\macaulay}{{\small\texttt{Macaulay}}\xspace}
\newcommand{\macaulaytwo}{{\small\texttt{Macaulay2}}\xspace}

\newcommand{\ie}{i.e.}
\newcommand{\eg}{e.g.}

\begin{document}

\title{\scsc: Satisfiability Checking meets\\ Symbolic Computation\\ (Project Paper)}
\author{
Erika \'Abrah\'am\inst{1} \and 
John Abbott\inst{12} \and
Bernd Becker\inst{2} \and
Anna M. Bigatti\inst{3} \and\newline
Martin Brain\inst{11} \and
Bruno Buchberger\inst{4} \and
Alessandro Cimatti\inst{5} \and\newline  
James H. Davenport\inst{6} \and
Matthew England\inst{7} \and
Pascal Fontaine\inst{9} \and\newline
Stephen Forrest\inst{10} \and
Alberto Griggio\inst{5} \and 
Daniel Kroening\inst{11} \and\newline
Werner M. Seiler\inst{12} \and
Thomas Sturm\inst{8,13}
}
\institute{ 
RWTH Aachen University, Aachen, Germany \and 
Albert-Ludwigs-Universit\"at, Freiburg, Germany \and
Universit\`a degli studi di Genova, Italy \and
Johannes Kepler Universit\"at, Linz, Austria \and
Fondazione Bruno Kessler, Trento, Italy \and
University of Bath, Bath, U.K. \and
Coventry University, Coventry, U.K. \and
CNRS, LORIA, Inria, Nancy, France \and
LORIA, Inria, Universit\'e de Lorraine, Nancy, France \and
Maplesoft Europe Ltd \and
University of Oxford, Oxford, U.K. \and
Universit\"at Kassel, Kassel, Germany \and
Max-Planck-Institut f\"ur Informatik, Saarbr\"ucken, Germany
}
\maketitle

\begin{abstract}

\emph{Symbolic Computation} and \emph{Satisfiability Checking} are two research areas, both having their individual scientific focus but sharing also common interests in the development, implementation and application of decision procedures for arithmetic theories. Despite their commonalities, the two communities are rather weakly connected. The aim of our newly accepted \scsc project (H2020-FETOPEN-CSA) is to strengthen the connection between these communities by creating common platforms, initiating interaction and exchange, identifying common challenges, and developing a common roadmap from theory along the way to tools and (industrial) applications.  In this paper we report on the aims and on the first activities of this project, and formalise some relevant challenges for the unified \scsc community.


\end{abstract}

\section{Introduction} 
\label{SEC:Intro}

The use of advanced methods to solve practical and industrially relevant problems by computers has a long history.  
While it is customary to think that ``computers are getting faster'' (and indeed, they were, and are still getting more powerful in terms of multicores etc.), the progress in algorithms and software has been even greater. One of the leaders in the field of linear and mixed integer programming points out \cite[slide 37]{Bixby2015a} that you would be over 400 times better off running today's algorithms and software on a 1991 computer than you would running 1991 software on today's computer. The practice is heavily inspired by the theory: \cite[slide 31]{Bixby2015a} shows that the biggest version-on-version performance advance in software was caused by ``mining the theory''.
\emph{But} this progress has been in what is, mathematically, quite a limited domain: that of linear programming, possibly where some of the variables are integer-valued. 

There has been also much progress in the use of computers to solve hard non-linear algebraic${}^{\ref{fn:AA}}$ problems. This is the area generally called \emph{Symbolic Computation} (or \emph{Computer Algebra}).  It includes solving non-linear problems over both the real and complex numbers, though generally with very different techniques. This has produced many new applications and surprising developments: in an area everyone believed was solved, non-linear solving over the reals (using cylindrical algebraic decomposition --- CAD) has recently found a new algorithm for computing square roots \cite{ErascuHong2014a}. CAD is another area where practice is (sometimes) well ahead of theory: the theory \cite{DavenportHeintz1988,BrownDavenport2007} states that the complexity is doubly exponential in the number of variables, but useful problems can still be solved in practice (\hspace{1sp}\cite{Araietal2014a} points out that CAD is the most significant engine in the ``Todai robot'' project).

Independently and contemporaneously, there has been a lot of practical progress in solving the SAT problem, \ie, checking the satisfiability of logical problems over the Boolean domain. The SAT problem is known to be NP-complete \cite{cook:np_complete}. 
Nevertheless, the \emph{Satisfiability Checking} \cite{handbook} community has developed SAT solvers which can successfully handle inputs with millions of
Boolean variables.  Among other industrial applications, these tools are now at
the heart of many techniques for verification and security of computer systems.


Driven by this success, big efforts were made to enrich propositional
SAT-solving with solver modules for different theories. Highly
interesting techniques were implemented in \emph{SAT-modulo-theories
  (SMT) solvers} \cite{Barrett14,decision_proc} for checking easier theories, but
the development for quantifier-free non-linear real and integer
arithmetic\footnote{It is usual in the SMT community to refer to these
  constraints as \emph{arithmetic}. But, as they involve quantities as
  yet unknown, manipulating them is \emph{algebra}.  Hence both words
  occur, with essentially the same meaning, throughout this
  document.\label{fn:AA}} is still in its infancy.

\definecolor{darkred}{rgb}{0.65,0,0}
\definecolor{darkblue}{rgb}{0,0,0.4}
\definecolor{owngreen}{rgb}{0,0.6,0}
\definecolor{darkgreen}{rgb}{0,0.35,0}
\definecolor{green}{rgb}{0,0.55,0}
\definecolor{gray}{rgb}{0.7,0.7,0.7}
\definecolor{darkgrey}{rgb}{0.7,0.7,0.7}
\definecolor{brown}{cmyk}{0, 0.8, 1, 0.6}

\newcommand{\cbb}[1]{{\color{blue}#1}}
\newcommand{\cdb}[1]{{\color{darkblue}#1}}
\newcommand{\cg}[1]{{\color{green}#1}}
\newcommand{\cdg}[1]{{\color{darkgreen}#1}}
\newcommand{\cred}[1]{{\color{red}#1}}
\newcommand{\cdred}[1]{{\color{darkred}#1}}
\newcommand{\cG}[1]{{\color{gray}#1}}

\begin{figure}[t]
\begin{center}
  \scalebox{0.85}{
    \begin{tikzpicture}
      \path[draw,thick,-latex'] (-0.1,0)--(12.3,0);
      \path[draw] (0,-0.1)--(0,0.1);
      \path[draw] (2,-0.1)--(2,0.1);
      \path[draw] (4,-0.1)--(4,0.1);
      \path[draw] (6,-0.1)--(6,0.1);
      \path[draw] (8,-0.1)--(8,0.1);
      \path[draw] (10,-0.1)--(10,0.1);
      \path[draw] (12,-0.1)--(12,0.1);
      \node at (0,-0.3) {\tiny 1960};
      \node at (2,-0.3) {\tiny 1970};
      \node at (4,-0.3) {\tiny 1980};
      \node at (6,-0.3) {\tiny 1990};
      \node at (8,-0.3) {\tiny 2000};
      \node at (10,-0.3) {\tiny 2010};
      \node at (12,-0.3) {\tiny 2020};
      
      \node at (0.5,2.5) {\large \cbb{CAS}};
      \node at (0.5,-2) {\large \cred{SAT}};
      \node at (0.5,-3) {\large \cg{SMT}};
      
      \node[rotate=60,anchor=west] at (0.6,0.25) {\scriptsize\tt\cbb{Schoonschip}};
      \node[rotate=60,anchor=west] at (1.0,0.25) {\scriptsize\tt\cbb{MATHLAB}};
      \node[rotate=60,anchor=west] at (1.6,0.25) {\scriptsize\tt\cbb{Reduce} \cdb{Altran}};
      \node[rotate=60,anchor=west] at (2.2,0.25) {\scriptsize\tt\cbb{Scratchpad/Axiom}};
      \node[rotate=60,anchor=west] at (3.4,0.25) {\scriptsize\tt\cbb{Macsyma}};
      \node[rotate=60,anchor=west] at (3.8,0.25) {\scriptsize\tt\cbb{SMP}};
      \node[rotate=60,anchor=west] at (4.2,0.25) {\scriptsize\tt\cbb{muMATH}};
      \node[rotate=60,anchor=west] at (4.6,0.25) {\scriptsize\tt\cbb{Maple}};
      \node[rotate=60,anchor=west] at (5.0,0.25) {\scriptsize\tt\cbb{Mathcad} \cdb{SAC} \cbb{GAP}};
      \node[rotate=60,anchor=west] at (5.4,0.25) {\scriptsize\tt\cbb{CoCoA} \cdb{MathHandbook} \cbb{Mathomatic}};
      \node[rotate=60,anchor=west] at (5.8,0.25) {\scriptsize\tt\cbb{Mathematica} \cdb{Derive} \cbb{FORM}};
      \node[rotate=60,anchor=west] at (6.2,0.25) {\scriptsize\tt\cbb{KASH/KANT} \cdb{PARI/GP}};
      \node[rotate=60,anchor=west] at (6.8,0.25) {\scriptsize\tt\cbb{Magma} \cdb{Fermat} \cbb{Erable} \cdb{Macaulay2}};
      \node[rotate=60,anchor=west] at (7.2,0.25) {\scriptsize\tt\cbb{Singular} \cdb{SymbolicC++}};
      \node[rotate=60,anchor=west] at (7.6,0.25) {\scriptsize\tt\cbb{Maxima}};
      \node[rotate=60,anchor=west] at (8,0.25) {\scriptsize\tt\cbb{Xcas/Giac}};
      \node[rotate=60,anchor=west] at (8.4,0.25) {\scriptsize\tt\cbb{Yacas}};
      \node[rotate=60,anchor=west] at (8.8,0.25) {\scriptsize\tt\cbb{SAGE} \cdb{SMath} \cbb{Studio}};
      \node[rotate=60,anchor=west] at (9.2,0.25) {\scriptsize\tt\cbb{Cadabra} \cdb{SymPy} \cbb{OpenAxiom}};
      \node[rotate=60,anchor=west] at (9.6,0.25) {\scriptsize\tt\cbb{MATLAB} \cdb{MuPAD}};
      \node[rotate=60,anchor=west] at (10.0,0.25) {\scriptsize\tt\cbb{Wolfram Alpha} \cdb{TI-Nspire} \cbb{CAS}};
      \node[rotate=60,anchor=west] at (10.5,0.25) {\scriptsize\tt\cbb{Mathics} \cdb{Symbolism} \cbb{FxSolver}};
      \node[rotate=60,anchor=west] at (11,0.25) {\scriptsize\tt\cbb{Calcinator} \cdb{SyMAT} \cbb{Mathemagix}};
    
    \node[rotate=-60,anchor=west] at (6.8,-0.4) {\scriptsize\tt\cred{WalkSAT} \cdred{SATO}};
    \node[rotate=-60,anchor=west] at (7.1,-0.4) {\scriptsize\tt\cg{Simplify} \cdg{SVC}};
    \node[rotate=-60,anchor=west] at (7.4,-0.4) {\scriptsize\tt\cred{GRASP} \cdred{Chaff} \cred{BCSAT}};
    \node[rotate=-60,anchor=west] at (7.8,-0.4) {\scriptsize\tt\cred{MiniSAT} \cdred{Berkmin} \cred{zChaff} \cdred{Siege}};
\node[rotate=-60,anchor=west] at (8.2,-0.4) {\scriptsize\tt \cg{ICS} \cdg{Uclid}  \cg{MathSAT} \cdg{Barcelogic} };
\node[rotate=-60,anchor=west] at (8.6,-0.4) {\scriptsize\tt\cred{HyperSat} \cdred{RSat} \cred{Sat4j}};
\node[rotate=-60,anchor=west] at (9.0,-0.4) {\scriptsize\tt\cg{Yices} \cdg{CVC} \cg{HySAT/iSAT} \cdg{DPT}};

\node[rotate=-60,anchor=west] at (9.3,-0.4) {\scriptsize\tt\cg{Z3} \cdg{Alt-Ergo} \cg{Beaver} \cdg{ABsolver} };
\node[rotate=-60,anchor=west] at (9.7,-0.4) {\scriptsize\tt\cg{Boolector} \cdg{PicoSAT} \cg{Spear}};
\node[rotate=-60,anchor=west] at (10.3,-0.4) {\scriptsize\tt\cg{MiniSmt} \cdg{STP} \cg{veriT} \cdg{OpenCog}};
\node[rotate=-60,anchor=west] at (10,-0.4) {\scriptsize\tt\cred{ArgoSat}};
\node[rotate=-60,anchor=west] at (10,-0.4) {\scriptsize\tt\phantom{ArgoSat} \cg{OpenSMT} \cdg{SatEEn} \cg{SWORD}};
\node[rotate=-60,anchor=west] at (10.6,-0.4) {\scriptsize\tt\cred{Glucose} \cdred{CryptoMiniSat}};
\node[rotate=-60,anchor=west] at (10.6,-0.4) {\scriptsize\tt\phantom{Glucose CryptoMiniSat} \cdg{SONOLAR}};
\node[rotate=-60,anchor=west] at (10.9,-0.4) {\scriptsize\tt\cred{Lingeling} \cdred{UBCSAT}};
\node[rotate=-60,anchor=west] at (10.9,-0.4) {\scriptsize\tt\phantom{Lingeling UBCSAT} \cg{SMTInterpol}};
\node[rotate=-60,anchor=west] at (11.2,-0.4) {\scriptsize\tt\cg{SMT-RAT} \cdg{SMCHR} \cg{UCLID} \cdg{Clasp}};
\node[rotate=-60,anchor=west] at (11.5,-0.4) {\scriptsize\tt\cred{Fast SAT Solver}};
\node[rotate=-60,anchor=west] at (11.5,-0.4) {\scriptsize\tt\phantom{Fast SAT Solver} \cg{raSAT}};

\end{tikzpicture}
}
\end{center}
\vskip-10pt
\caption{History of some computer algebra systems and SAT/SMT solvers (not exhaustive; years approximate first release as far as known and as positioning allowed) \protect\cite{Abraham2015b}}
\vskip-10pt
\label{fig:tools}
\end{figure}

Figure \ref{fig:tools} shows a non-exhaustive history of tool developments in these two areas. It illustrates nicely the historically deeper roots of computer algebra systems, but also the high intensity of research in both areas. The resulting tools are successfully applied in several academic and industrial areas, however, the current state is still not satisfactory, as described in \cite{Platzeretal2009}: 

\begin{quotation}\noindent
``\emph{Despite substantial advances in verification technology, complexity issues with classical decision procedures are still a major obstacle for formal verification of real-world applications, \eg, in automotive and avionic industries.}''
\end{quotation}
Both communities address similar problems and share the challenge to
improve their solutions to achieve applicability on complex
large-scale applications.  However, the Symbolic Computation community
and the Satisfiability Checking community are largely in their own
silos and traditionally do not interact much with each other.

To connect these communities, we successfully applied
for a European Horizon 2020 \emph{Coordination and Support Action},
 with an envisaged project start in July 2016.  The
overall aim of this project is to create a new research community
bridging the gap between Satisfiability Checking and Symbolic
Computation, whose members will ultimately be well informed about both
fields, and thus able to combine the knowledge and techniques of both
fields to develop new research and to resolve problems (both academic
and industrial) currently beyond the scope of either individual field.
We call the new community \scsc,
as it will join the communities of \textbf{S}atisfiability \textbf{C}hecking and
\textbf{S}ymbolic \textbf{C}omputation.

The contributions of this paper are twofold: Firstly, we discuss the potentials of closer connection and more intensive exchange between the two communities, and list a number of challenges that are currently out of reach but could be tackled by a unified \scsc community (Section \ref{SEC:Challenges}). Secondly, we discuss what is needed to trigger and support these developments, and describe the actions of our project to satisfy these needs (Section \ref{SEC:Actions}).



\section{Background}
\label{SEC:Background}

Before describing our project,
we give a short description of the state-of-the-art in Satisfiability
Checking and Symbolic Computation. Parts of this section are taken
from \cite{Abraham2015b}.

\subsection{Symbolic Computation and Computer Algebra Systems}

Computer Algebra, the use of computers to do algebra rather than simply
arithmetic, is almost as old as computing itself, with the first PhD theses
\cite{Kahrimanian1953,Nolan1953}
dating back to 1953.
%
This initial work  consisted of programs to
do one thing, but the focus soon moved on to `systems', capable of doing a
variety of tasks. One early such system was Collins' \sac
\cite{Collins1971b}, written in Fortran. Many of the early systems were written in LISP, largely because of
its support for recursion, garbage collection and large integers. The group at M.I.T. developed \macsyma~\cite{MartinFateman1971} in the 1960s.
about the same time, Hearn developed \reduce \cite{Hearn2005}, and
shortly after a group at IBM Yorktown Heights produced \scratchpad,
then \axiom \cite{JenksSutor1992}, a system that attempted to match
the generality of Mathematics with some kind of generic programming,
to allow algorithms to be programmed in the generality in which they
are conventionally stated, \eg, polynomials over a ring.

Symbolic Computation was initially seen as part of Artificial Intelligence, with major triumphs such as \cite{Slagle1961} being ``A Heuristic Program that Solves Symbolic Integration Problems in Freshman Calculus'', firmly in the AI camp. By the end of the 1960s, this approach to integration had been replaced by an algorithm \cite{Moses1967}, which had the great advantage that, when backed up with a suitable completeness theorem \cite{Risch1969a} it could \emph{prove} unintegrability: ``there is no formula made up of exponentials, logarithms and algebraic functions which differentiates to $e^{-x^2}$'', in other words ``$e^{-x^2}$ is unintegrable''.

The 1960s and 70s also saw great advances in other areas. We had much more efficient algorithms to replace naive use of Euclid's algorithm for greatest common divisor computation (and hence the simplification of fractions), far better algorithms than the search-based methods for polynomial factorisation, and so on. All this meant that Symbolic Computation firmly moved into the camps of algorithmics and complexity theory, and the dominant question became ``what is the worst-case complexity of this algorithm''.

\medskip\noindent\emph{Gr\"obner bases.}\quad
One great success of this period was the method of \emph{Gr\"obner bases} \cite{Buchberger1965}. This allows effective, and in many cases efficient, solution of many problems of polynomials over algebraically-closed fields (typically the complex numbers, though applications over finite fields and in cryptography abound).
This notion paved the way for the discovery of numerous effective
methods for polynomial ideals; many applications in other areas of
Mathematics quickly followed.
Buchberger's algorithm for computing a Gr\"obner basis is a prime
example of the huge gulf that can separate an abstract algorithm from
a usably efficient implementation. Over the fifty years since its
initial publication, research into the algorithm's behaviour has
produced several significant improvements: the modern refined
version is typically thousands of times faster than the original. The
search for further improvements continues today.

The remarkable computational utility of Gr\"obner bases prompted the
development of a number of distinct, independent implementations of
refined versions of Buchberger's algorithm. The main commercial
general-purpose 
computer algebra systems (including \magma~\cite{magma}, \maple~\cite{maple}, \mathematica~\cite{mathematica}) can all
compute Gr\"obner bases;
researchers needing the flexibility and ability to experiment with new algorithms also use computer algebra systems such as
\cocoa/\cocoalib~\cite{CoCoA-5},
\macaulay/\macaulaytwo \cite{M2} and 
\singular \cite{Singular} and
\reduce~\cite{Hearn2005}
which are freely 
downloadable from their respective websites.


\medskip\noindent\emph{Cylindrical algebraic decomposition.}\quad
%
Another great success of the 1970s was the development of \emph{cylindrical algebraic decomposition} (\emph{CAD}) in \cite{Article_Collins_75}. This replaced the non-elementary complexity (no finite tower of exponentials bounds the complexity) of Tarski's method for real algebraic geometry, by a doubly exponential method. A CAD is a decomposition of $\mathbb{R}^n$ into cells arranged cylindrically (meaning their projections are equal or disjoint) and described by semi-algebraic sets. 
For a detailed description of modern CAD, see \cite{Bradfordetal2016a}.
Hong created a
 C version of both the \texttt{SAC} library and the comprehensive CAD code, which
 is now open-source and freely available as \texttt{SACLIB} and
 \texttt{QEPCAD-B} \cite{Article_Brown_QEPCAD}.
 Another example is the \redlog
 package~\cite{DolzmannSturm:97a} of the computer algebra system
 \reduce, which offers an optimised combination of the cylindrical algebraic
 decomposition with virtual substitution (see below) and Gr\"obner basis methods.

\medskip\noindent\emph{Virtual substitution.}\quad
To mention a last algorithm,
\emph{virtual substitution}
\cite{Article_Weispfenning_Quadratic} focuses on non-linear real
arithmetic formulas where the degree of the quantified variables is
not too large. Although the method can be generalised to arbitrary
degrees, current implementations are typically limited to input, where
the total degree of the quantified variables does not exceed $2$. In
practice, this limitation is somewhat softened by employing powerful
heuristics like systematic \emph{degree shifts} or polynomial
factorisation. One key idea is to eliminate existential quantifiers in
favour of \emph{finite} disjunctions plugging in test terms that are
derived from the considered formula.

These methods and their numerous refinements belong to the usual tool
box of state-of-the-art computer algebra systems, and enable them to
tackle hard arithmetic problems. 

\subsection{Satisfiability Checking}



In the 1960s, another line of research on \emph{Satisfiability
  Checking} \cite{handbook} for \emph{propositional logic} started its career. The
first idea used \emph{resolution} for quantifier
elimination~\cite{DP}, and had serious problems with the steeply increasing requirements on computational and memory resources with the increase of the problem size. Another research
line~\cite{DLL} suggested a combination of \emph{enumeration} and
\emph{Boolean constraint propagation (BCP)}. A major improvement was
achieved in the 1990s by \emph{combining} the two approaches, leading
to \emph{conflict-driven clause-learning} and \emph{non-chronological backtracking} \cite{CDCL}. Later on, this
impressive progress was continued by novel efficient implementation
techniques (\eg, sophisticated decision heuristics, two-watched-literal
scheme, restarts, cache performance, etc.), resulting in numerous
powerful \emph{SAT solvers}. 

Driven by this success, big efforts were made to enrich propositional
SAT-solving with solver modules for different existentially quantified theories. Highly
interesting techniques were implemented in \emph{SAT-modulo-theories
  (SMT) solvers} for checking, \eg, equality logic with uninterpreted
functions, array theory, bit-vector arithmetic and quantifier-free
linear real and integer arithmetic, but the development
for quanti\-fier-free non-linear real and integer arithmetic 
is still in its infancy. 
For further reading, see, \eg,
\cite{Barrett14,decision_proc}.



\begin{figure}[t]
\begin{center}
\scalebox{0.8}{
\begin{tikzpicture}[font=\large\bfseries, auto, thick, 
box/.style={draw, rounded corners=3pt, inner sep=5pt},
boxA/.style={draw, rectangle split,rectangle split parts=2,text centered,rounded corners=3pt, inner sep=5pt},
boxB/.style={draw, rectangle split, rectangle split horizontal=false,rectangle split parts=2,text centered, rounded corners=3pt, inner sep=5pt}]

\node [box] (ssolver) [color=blue] at (0,0) {SAT solver};
\node [box,color=red] (phi) [above=1.2 of ssolver] {\small input formula in CNF};
\node [box] (eq) [color=blue] at (-3,-1.5) {\small theory constraint set};
\node [box] (ex) [color=blue] at (3,-1.5) {\small \begin{tabular}{c}(partial) SAT or \\ UNSAT $\small \color{blue}+$ \color{blue}explanation\end{tabular}};
\node [box] (tsolver) [color=blue] at (0,-3) {theory solver(s)};

\node [box] (UNSAT) [right=4 of ssolver,color=red] {\small \begin{tabular}{c}SAT or\\UNSAT\end{tabular}};

\draw [arrows={-triangle 60},color=red] (ssolver) edge node [above] {\small solution or} node[below] {\small unsatisfiable} (UNSAT);

\draw [arrows={-triangle 60},color=red] (phi) edge node [right, align=left] {\small Boolean abstraction} (ssolver);

\draw [color=blue,-] (ssolver) edge [bend right,out=-20] 
node [above left, xshift=0.8cm,yshift=-0.4cm,fill=white] {\small (partial) solution} 
(eq);
\draw [color=blue,arrows={-triangle 60}] (eq) edge [bend right] (tsolver);

\draw [color=blue,-] (tsolver) edge [bend right] 
(ex);
\draw [color=blue,arrows={-triangle 60}] (ex) edge [bend right,in=200] (ssolver);

\end{tikzpicture}
}
\end{center}
\vskip-10pt
\caption{The functioning of SMT solvers}
\vskip-10pt
\label{fig:smt}
\end{figure}
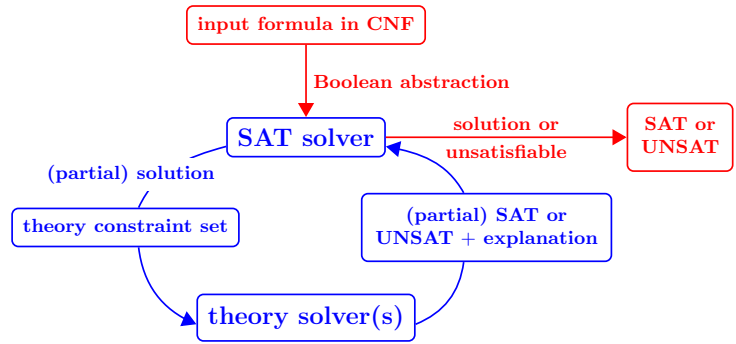


Modern \emph{SMT solvers} typically combine a \emph{SAT solver} with
one or more \emph{theory solvers} as illustrated in
Figure~\ref{fig:smt}.  First the input formula is transformed into
conjunctive normal form (CNF), a conjunction of disjunctions (clauses);
this transformation can be done in linear time and space using
Tseitin's transformation on the cost of additional variables.  Next,
the resulting CNF is abstracted to a pure Boolean propositional logic formula
by replacing each theory constraint by a fresh Boole\-an proposition.
Intuitively, the truth value of each fresh proposition defines whether
the theory constraint, which it substitutes, holds.  The SAT solver
tries to find solutions for this propositional abstraction and during
solving it consults the theory solver(s) to check the consistency of
the theory constraints that should hold according to the current
values of the abstraction variables.

On the one hand, theory solvers only need to check
\emph{conjunctions (sets)} of theory constraints, instead of arbitrary
Boolean combinations. On the other hand, theory solvers should have
the following properties for being \emph{SMT-compliant}:
\begin{itemize}
\item They should work \emph{incrementally}, \ie, after
they determine the consistency of a constraint set, they should be
able to take delivery of some additional constraints and re-check the
extended set, thereby making use of results from the previous check.
\item In case of unsatisfiability, they should be able to return an
  \emph{explanation} for inconsistency, \eg, by a preferably
  small inconsistent subset of the constraints.
\item They should support \emph{backtracking}, \ie, the removal of
  previously added constraints.
\end{itemize}
Optimally, theory solvers should also be able to provide  a \emph{satisfying solution}, if the problem is satisfiable, and a \emph{proof of unsatisfiability} for the explanation, if the problem is unsatisfiable.

A great advantage of the SMT technology is that it can employ decision
procedures not only in isolation, but also \emph{in combination}.  For
example, solving non-linear arithmetic formulas can often be speeded
up by first checking linear abstractions or linear problem parts using
more efficient decision procedures, before applying heavier
procedures. Additionally, theories can also be combined already in the
input language of SMT solvers.  For example, deductive program
verification techniques generate verification conditions, which might
refer to arrays, bit-vectors as well as integers; in such cases,
dedicated SMT solvers can apply several decision procedures for
different theories in combination.

When combining decision procedures, \emph{incomplete} but
\emph{efficient} procedures are also valuable, if they
guarantee termination but not necessarily return a conclusive answer.
Such incomplete methods are frequently applied in
SMT solving, a typical example being interval constraint propagation,
based on interval arithmetic. Some solvers combine such incomplete
methods with complete decision procedures, in order to guarantee the
solution of the problem, while increasing efficiency. Other solvers
even sacrifice completeness and might return a ``don't know'' answer,
but still they are able to solve certain extremely large problems,
which are out of reach for complete methods, very fast. Furthermore,
incomplete procedures are the only way to support problems from
undecidable theories, like formulas containing exponential or
trigonometric functions.

SAT and SMT solvers are tuned for efficiency. Combining complete and
incomplete decision procedures, making use of efficient heuristics,
learning not only propositional facts but also (Boolean abstractions
of) theory lemmas at the SAT level allow modern SMT solvers to solve
relevant large-size problems with tens of thousands of variables,
which could not be solved before by single decision procedures in
isolation.  For some example applications see, \eg,
\cite{problem_solving}.

Examples for solvers that are able to cope with linear arithmetic problems (either in a
complete or in an incomplete manner) are \altergo \cite{altergo},
\cvc~\cite{cvc4}, \isat~\cite{Article_Fraenzle_HySAT07,ScheiblerKB13},
\mathsat~\cite{mathsat}, \opensmt \cite{Article_Bruttomesso_OpenSMT},
\smtrat \cite{smt-rat}, \verit~\cite{Bouton1}, \yices~\cite{DM06}, and
\zthree~\cite{z3}. A further interesting SMT-approach for linear integer arithmetic is
proposed in \cite{sturm*:cutsat}.

Much less activity can be observed for SMT solvers for non-linear
arithmetic. A few SMT tools embedded some (complete as well as
incomplete) decision procedures.
Such a solver is \isat, which
uses interval constraint propagation.  The SMT solver
\minismt~\cite{zankl*:minismt} tries to reduce non-linear real arithmetic
problems to linear real arithmetic and can solve only satisfiable
instances this way.  We are aware of only two SMT solvers that are
complete for non-linear real arithmetic: Firstly, the prominent
\zthree solver developed at Microsoft Research, which uses an elegant
SMT-adaptation of the cylindrical algebraic decomposition
method~\cite{nlsat}.  Secondly, 
\smtrat~\cite{smt-rat},
using solver modules for simplex, the
cylindrical algebraic decomposition,
the virtual substitution method, Gr\"obner
bases, interval constraint propagation,
branch and bound, and their strategic combination~\cite{deMoura_strategy}.

Even fewer SMT solvers are available for non-linear integer arithmetic, which is undecidable in
general. A linearisation approach was proposed in
\cite{BorEtAlNiaToLia09}.
The SMT solving spin-off of \aprove~\cite{codish2012exotic}
uses bit-blasting.  To our knowledge, \zthree implements a combination of linearisation
and bit-blasting. \isat uses interval constraint propagation, whereas
\altergo combines the idea of~\cite{Bobot_asimplex-based} with an
axiom-based version of interval constraint propagation.  
\smtrat can tackle this theory using a generalised
branch-and-bound technique.

\smallskip

The increasing variety of the theories considered
by SMT solvers created an urgent need for a common input language.  The
\smtlib initiative \cite{SMTLIB-url} defined a \emph{standard
  input language} for SMT solvers with a first release in 2004, and
provides a large and still increasing number of \emph{benchmarks},
systematically collected for all supported theories. \smtlib
also enabled the start of \emph{SMT competitions}; the first one took
place in 2005 with 12 participating solvers in 7 divisions (theories,
theory combinations, or fragments thereof) on 1360 benchmarks, which
increased in 2014 to 20 solvers competing in 32 divisions on 67426
benchmarks.
The \smtlib standard and the competitions not only intensified the SMT
research activities, but also gave visibility and acceptance for SMT
solving in computer science and beyond. Once a problem is formulated
in the \smtlib language, the user can employ
\emph{any} SMT solver to solve the problem.

\section{Some Scientific Challenges and Opportunities}
\label{SEC:Challenges}

On the one hand, SMT solving has its strength
in efficient techniques for exploring Boolean structures, learning,
combining solving techniques, and developing dedicated heuristics, but
its current focus lies on easier theories and it makes use of Symbolic
Computation results only in a rather naive way. There are fast SMT
solvers available for the satisfiability checking of linear real and
integer arithmetic problems, but just a few can handle non-linear
arithmetic.
On the other hand, Symbolic Computation is strong in providing
powerful procedures for sets (conjunctions) of arithmetic constraints,
but it does not exploit the achievements in SMT solving for
efficiently handling logical fragments, using heuristics and learning
to speed-up the search for satisfying solutions.

The Satisfiability Checking community would definitely profit from further
exploiting Symbolic Computation achievements and adapting and extending them
to comply with the requirements on embedding in the SMT context.
However, it is a highly challenging task, as it requires a deep
understanding of complex mathematical problems, whose embedding in SMT
solving is not trivial.

Symmetrically, Symbolic Computation could profit from exploiting
successful SMT ideas, but it requires expertise in efficient solver
technologies and their implementation, like dedicated data structures,
sophisticated heuristics, effective learning techniques, and
approaches for incrementality and explanation generation in theory
solving modules. 

In this section we describe some ideas of how algorithms and tools from
both communities could be made more powerful by exploiting scientific
exchange and technology transfer.

\subsection{Symbolic Computation Techniques for Satisfiability Checking}

Many practical decision procedures, designed by the Symbolic
Computation community, are implemented in computer algebra systems
(\eg, linear real and integer arithmetic, non-linear real
arithmetic, linear programming, quantified formulas, Gr\"obner and
involutive bases).  To use them in a Satisfiability Checking context,
some scientific and engineering issues need solutions, notably to find
new ways of \emph{incremental} solving, \emph{explaining}
unsatisfiability and generating \emph{lemmas}.

Whereas for linear real arithmetic useful procedures have been adapted
to satisfy the requirements for SMT embedding, many opportunities
remain to be explored for non-linear arithmetic.  For example, there
are (to the best of our knowledge) just two SMT solvers, \zthree and
\smtrat, which make use of the CAD method, but in a different way:
\zthree uses a very elegant solution to explore the state space by a
close integration of theory decisions and theory propagation in the
Boolean SAT search, and constructs CAD only partially to explain
conflicts in the above search (more precisely, to compute a
semi-algebraic description of CAD cells that do not satisfy a given
sign condition). In contrast, \smtrat implements an incremental
version of the CAD method, which works hand-in-hand with the search at
the logical level. For this latter approach, the power of heuristics
(variable and polynomial ordering for incremental projection, choice
and order of sample points for lifting) and the generation of lemmas
(most importantly the computation of explanations for
unsatisfiability) is still far from being fully exploited.

There is still great potential for improvements not only for the
SMT embedding of the CAD method, but also other non-linear arithmetic
decision procedures like virtual substitution or Gr\"obner
bases, and their strategic combination with each other and further
light-weight methods such as interval constraint propagation.

Another important aspect is the Symbolic Computation community's
expertise in simplification and preprocessing.  The complexity of the
problems this community handles is often extremely high, and no
practical procedure would exist without significant techniques to
prepare the input problems.  Such techniques do exist for
Satisfiability Checking, but they rather focus on easier theories. A
transfer of the savoir-faire in simplification and preprocessing for
non-linear real and integer arithmetic would certainly be highly
profitable.

\subsection{Satisfiability Checking Techniques for Symbolic Computation}

A key ingredient in the success of the Satisfiability Checking tools
is the use of \emph{learning} and \emph{non-chronological
  backtracking} techniques to speed up the search through tree-shaped
search structures.  Traditionally (and in the majority of cases) CAD
proceeds through a two stage process: first, projecting the problem through
lower dimensions; then lifting: incrementally building a solution
in increasing dimensions.  An alternative approach using
triangular decomposition was introduced in \cite{Chenetal2009d} where
first the complex domain is cylindrically decomposed and then refined
to a CAD of the real domain, where all data is in a tree-shaped
structure.

Other techniques are certainly amenable for learning with
the non-chrono\-logical backtracking approach.  For instance, first
prototypes integrating CDCL-style learning techniques with virtual
substitution for linear quantifier elimination have been successfully
created and studied.  Integration of learning techniques with the
computation of comprehensive Gr\"obner bases \cite{Weispfenning:92a}
should also be investigated.

Incrementality, which played an important role in the success of
Satisfiability Checking, may also be used to make Symbolic Computation
techniques more efficient. The alternative CAD construction method
described above is also incremental in nature \cite{Bradfordetal2014b}
and so may offer one option here. An incremental CAD-based decision procedure
for solving polymial constraint systems was proposed in
\cite{Strzebonski:2012}. There exist algorithms for computing
Gr\"obner bases which exploit known mathematical facts about the ideal
generated by the basis like its Hilbert function or some syzygies.
Traditionally, this has been seen as a way to speed up computations.
However, these approaches can naturally be adapted into incremental
algorithms.

A central aspect of Satisfiability Modulo Theories is the combination
frameworks for theories and decision procedures.  Combining theories
in Symbolic Computation (combined real/floating point theories,
interval constraint propagation with other arithmetic theories) might
also bring a number of advantages and possibilities, for instance,
more expressive languages, or efficiency due to hierarchical
reasoning.  While combination of theories in Symbolic Computation are
typically very specific and ad hoc, 
the
SMT community systematically uses the generic Nelson--Oppen
framework~\cite{Nelson3}
for disjoint theories.  Such a
framework can of course not be used as it is, but it might be an
inspiration for a modular approach in Symbolic Computation.

\subsection{Standard Languages and Benchmarks}

The initiation and maintenance of a common problem specification
language standard \smtlib \cite{SMTLIB-url} and of competitions form
an important part of the Satisfiability Checking community effort.
Besides providing a stimulating event for tool developers to exhibit
their systems, the competitions are also a vehicle for publishing practical
progress.  Competition results are advertised, and consulted by users
to pick the best tools and techniques to solve problems.

The Symbolic Computation community does not have a similar tradition,
and indeed, to quote one major system developer: ``it is very hard to
get any practical improvements published~---~the reviewers will often
say this is not hard science''.  Although it is not good to only focus
on a small library of benchmarks and have the competition as sole goal,
competitions do have a tremendously positive effect on tools and
techniques, as witnessed in the Satisfiability Checking community,
especially if the competition challenges are concrete industrial
challenges. Such driving forces could be also established in Symbolic Computation.

Though in Satisfiability Checking the standard input language allowed
to provide large benchmark sets to the community, benchmarks for
non-linear arithmetic theories are still rare, and harder to describe
without ambiguity. Therefore, also the Satisfiability Checking
community would profit from a common standard with an increased number
of non-linear arithmetic benchmarks.

\section{Project Actions}
\label{SEC:Actions}

The solution of challenging problems, as mentioned in the previous
section, could be within reach, when supported by a stronger
collaboration between both \scsc research areas, creating an
infrastructure for dialogue and knowledge transfer.  However, the
research areas of Satisfiability Checking and Symbolic Computation are
still quite disconnected, as reflected in their communication
platforms and support structures.  Symbolic Computation has its own
conferences (ACA, CASC, ISSAC, etc.), several dedicated journals (\eg,
AAECC, JSC, MSC), and the SIGSAM forum.  Similarly, Satisfiability
Checking is supported by its own conferences (CADE, IJCAR, SMT, etc.)
and journals (\eg, JAR), the SatLive forum to keep up-to-date with research, SMT
standards, and SAT- and SMT-solver competitions.

The main aims of our project are to create \emph{communication
  platforms} and propose \emph{standards} to enable the interaction
between the two communities, and to use these platforms to \emph{initiate discussions
  and cooperation} and to \emph{identify potentials, challenges and
  obstacles} for future research and practical applications. In the following we shortly describe
planned actions of our \scsc project to achieve these goals.

\smallskip
\noindent \emph{Communication platforms}\quad To bridge the \scsc
communities, we will initiate platforms to support the interaction of
the currently disjoint groups. We organised a Dagstuhl Seminar \emph{Symbolic Computation and Satisfiability Checking}\footnote{\url{http://www.dagstuhl.de/en/program/calendar/semhp/?semnr=15471}}
15-20th November, 2015, which already led to numerous interesting
discussions and interactions.  At CASC 2016, we will organise a
\emph{topical session} devoted to topics from the cross-community
\scsc area.  Furthermore, we will establish a workshop series in the
area of \scsc, covering the interests of both communities, and having
its first edition affiliated with SYNASC 2016.  These workshops will
serve as platforms for scientific exchange, discussion and cooperation
within and between the currently disjoint communities.  To support and
attract young new community members, we will organise a dedicated
summer school aimed at interested young researchers from \scsc areas,
with courses specifically tailored to their needs.

\smallskip
\noindent \emph{Research roadmap} The above platforms will initiate 
cross-community interactions, and help to clearly identify unused potentials.
We aim at initiating discussions on what the
communities can learn from each other, what are the common challenges
which they can solve together, what Satisfiability Checking could
learn from Symbolic Computation achievements, and which Satisfiability
Checking results could be adapted to improve Symbolic Computation
solutions.

Our long-term objective is to create a research roadmap of potentials and
challenges, both to the two traditional subject silos, but also challenges that only the
new joined \scsc community can address.  This roadmap should identify,
within the problems currently faced in the industry, the particular
points that can be expected to be solved by the \scsc community in the
short and middle term, and will provide recommendations for spin-off
projects.

\smallskip
\noindent \emph{Standards, benchmarks, competitions} We aim to create
a standard problem specification language capable of representing
common problems of the \scsc community.  We plan on extending the
\smtlib language, which is already mature and fully accepted among the
SMT (Satisfiability Checking) community, to handle features needed for
the Symbolic Computation community.  This will be done in a modular
way, with a particular focus on extensibility for new features.

Agreeing on a common language, and being able to share challenging
problems is an essential aspect for building a dynamic community.
This will foster further discussions and uncover problems that can be
solved by the \scsc community altogether, set clear challenges on
which various approaches can be evaluated, classify the approaches
according to their strength and weaknesses on the various kinds of
problems.  Mixed approaches will naturally emerge, to tackle problems
exhibiting several orthogonal difficulties.  The standard could also
serve as a communication protocol for platforms mixing tools, to build
meta-tools to solve large and difficult problems out of reach of
current techniques often specialised to just one kind of job.

\smallskip
\noindent \emph{How to become an associate?} This project cannot reach its
aims by involving just a small number of core project members. To be
able to cover sufficiently wide research and application areas and to
take into account their needs and interests, there are currently $37$
\scsc \emph{associates} from both research communities as well as from industry. Our
associates will be regularly informed about the project activities and
they will be invited to take part in the corresponding events.


  The \scsc Coordination and Support Action will be an
  optimal platform for industrial and academic partners and associates
  to form smaller working groups and initiate specific projects.  If
  you would like to participate in the project as an associate,
  please contact the Project Coordinator James Davenport\footnote{Email
    contact: J.H.Davenport@bath.ac.uk}.

\section{Conclusions and Future Work} 
\label{SEC:Conclusion}

In this paper we gave a short description of the aims and actions of
our upcoming EU Coordination and Support Action \scsc.

The \scsc
project will maintain a website (\url{http://www.sc-square.org})
making readily accessible all the public information of the project
(\eg, contact information, details of past and forthcoming \scsc
workshops and other similar events).

\subsection*{Acknowledgements}

We thank the anonymous reviewers for their comments.
We are grateful for support by the H2020-FETOPEN-2016-2017-CSA project
\scsc (712689) and the ANR project ANR-13-IS02-0001-01 SMArT.  
Earlier work in this area was also supported by the EPSRC grant EP/J003247/1. 

\bibliography{literature}

\end{document}